\documentclass{elsart}
\usepackage{psfig,natbib}
\begin{document}
\runauthor{Goodrich}
\begin{frontmatter}

\title{FOS Observations of Four NLS1s}
\author[Keck]{Robert W. Goodrich}
\address[Keck]{The W. M. Keck Observatory, Kamuela, HI}

\thanks[HST]{This work has been supported by NASA
grant GO-06766.01-95A.}

\begin{abstract}
Ultraviolet through optical spectroscopy of four NLS1s shows strong
absorption features in the high-ionization UV resonance lines
such as C~IV, N~V, and Si~IV.
Mg~II is not absorbed.  The absorption could originate in the 
warm absorber.
\end{abstract}
\end{frontmatter}

\section{Introduction}
We observed four NLS1s with the HST   
Faint Object Spectrograph (FOS).
Spectra were obtained on consecutive orbits to provide near-simultaneous
coverage from the UV through the optical ($\lambda\lambda1150-6800$). 
The effects of variability should be minimized in these data.  
Three of the targets,
Mrk 493, WPVS007, and RX J$0134-42$, show strong optical Fe II emission
and weak [O III] $\lambda5007$ narrow-line emission.  Two objects,
WPVS007 and RX J$0134-42$, show steep X-ray spectra and large X-ray
variability.  Two, Akn 564 and RX J$0134-42$, have been reported to
have warm X-ray absorbers.

The original goals were: (a) to calculate physical parameters of the
gas from the diagnostic UV emission lines, (b) to study the UV iron-line
complexes, and (c) to obtain a good piece of the spectral energy distribution
(SED).

\section{UV Absorption}

We found strong absorption in the UV resonance lines in three of the
four NLS1s: Akn 564, WPVS007, and RX J$0134-42$.  In the fourth, Mrk 493,
asymmetric UV line profiles may hint at weak absorption on the red side of
the line (discussed below).  
The absorption is seen at low velocities in high-ionization 
lines such as Ly$\alpha$, N~V, Si~IV, and C~IV, but not in
lower-ionization lines such as Mg~II $\lambda\lambda2796,2804$.

\subsection{Akn 564}

Figure 1 shows the Ly$\alpha$/N V region of the FOS spectrum.  The zero-point
of the velocity scale was determined from the H$\beta$ emission line, which
does not show absorption.  Note the strong, zero-velocity absorption in both
Ly$\alpha$ and the N V $\lambda\lambda1239,1243$ doublet.  In the latter
case there is some evidence that the absorption extends below the local 
continuum level, but this should be confirmed with higher resolution, higher
S/N spectroscopy.

There are two observational effects of such absorption.  First, a spectrum
with poorer S/N and/or poorer resolution may not resolve the absorption
feature.  However, the resulting line profile would look broader than
the true emission line profile, since the absorption is removing flux
near the line center.  Secondly, it becomes highly problematic to
determine the original emission line flux, and hence to determine
physical parameters (density, ionization parameter, etc.) of the
emission-line gas from the UV resonance lines.

A footnote about this region of the spectrum.  At $-4100$ km s$^{-1}$ there
is an apparent, weak emission line with central absorption.  The 
combination of emission and absorption makes this identification as
a weak line reasonably secure, and the sharpness of the absorption
feature allows us to pinpoint the wavelength: $1198.95~\pm~0.25$ {\AA}.
A search of wavelengths of known high ionization resonance lines near this
wavelength indicates that this may be S V] $\lambda1199.134$, although
this identification must be considered tentative.

\begin{figure}
\centerline{\psfig{file=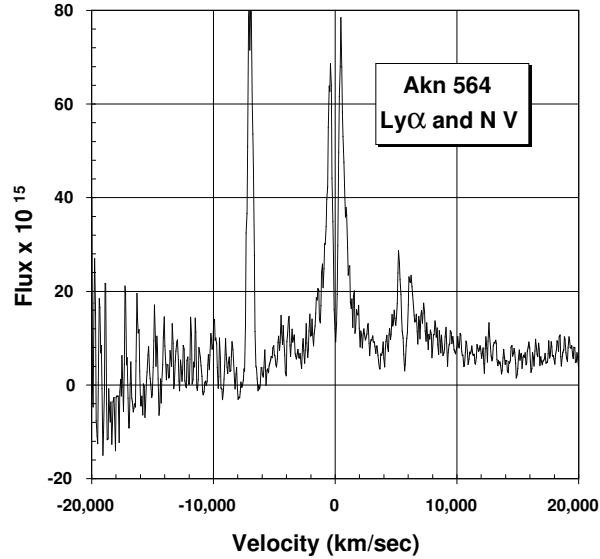,width=8cm,clip=}}
\caption{Strong absorption in the UV resonance lines of Akn 564
is clearly visible.  Note the absorption in both Ly$\alpha$ (at
zero velocity)
and the adjacent N V $\lambda1240$ doublet.}
\label{fig1}
\end{figure}

\subsection{WPVS007}

WPVS007 (Fig. 2) shows absorption similar to that seen in Akn 564.
It is slightly blue-shifted, with a second component seen 
near $-2500$~km~s$^{-1}$.

\begin{figure}
\centerline{\psfig{file=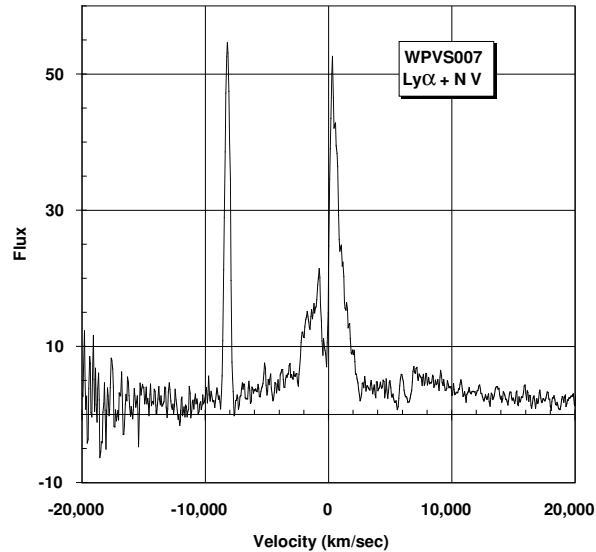,width=8cm,clip=}}
\caption{The UV absorption in WPVS007 is more complex than in Akn 564.}
\label{fig2}
\end{figure}

\subsection{RX J0134--42}

RX J$0134-42$ is unusual, showing very weak emission lines and apparently
red-shifted absorption.  It is the faintest of the four NLS1s observed,
and our spectrum (Fig. 3) is of lower quality.
There is clear
absorption in the C~IV line, and the N~V line profile (not shown)
is consistent with somewhat weaker absorption.  It is worth
noting that emission seen at $\lambda1232$ is most
likely blue-shifted N~V
emission, corresponding closely to the blue-shifted C~IV emission peak seen in
Fig. 3.  There is apparently no sign of Ly$\alpha$ emission!

\begin{figure}
\centerline{\psfig{file=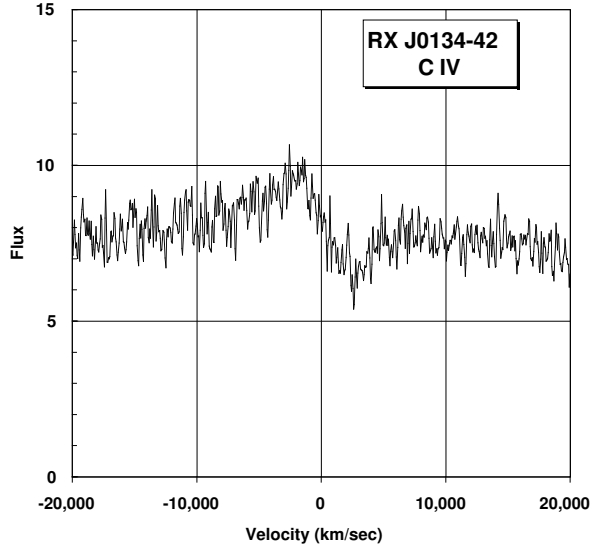,width=8cm,clip=}}
\caption{Strong red-shifted absorption is seen in the C IV line of 
RX J$0134-42$.  
The zero-point of the velocity scale corresponds to the red C IV
line at the redshift determined from the unabsorbed H$\beta$ emission line.}
\label{fig3}
\end{figure}

Red-shifted absorption is rather unusual in AGN.
Blue-shifted absorption from outflows lying in front of the AGN's
continuum source is common and sometimes occurs at very high
velocities (e.g. broad-absorption line QSOs), but
red-shifted absorption is rare and generally occurs only at low velocities.
The occurrence of strong red-shifted absorption in RX J$0134-42$
supports the interpretation of the blue asymmetries seen
in some NLS1 UV emission lines (e.g. Mrk 493, not shown in this 
contribution) as being formed by weaker red absorption, rather than a broad
blue wing from the emission source.

\subsection{Discussion}

Note that the high incidence of UV absorption in our sample
differs from that of the Leighly sample presented at this conference.
Small number statistics could be at work, or there could be some
difference in the selection criteria.  Luminosity has been suggested
as one possible difference.

\section{Relationship to the Warm Absorber}

X-ray spectra of Akn 564 were presented at this conference which were
variously interpreted as showing absorption at 1 keV, emission at
1 keV, or a lack of any features.  This may be a result of X-ray 
variability, but clearly higher spectral resolution and signal-to-noise
is needed in order to derive physical parameters such as the ionization
level.  This can be compared to the ionization level apparent in the
UV absorber.  Currently there seems to be no compelling evidence that
to conclude that the two absorbers are different.  Given the large
X-ray variability in some NLS1s, it seems reasonable to obtain
multi-epoch UV observations to search for variability of both the emission
and absorption features seen in our FOS data.

The argument has been made that the X-ray absorber cannot be outflowing
at the high velocities typical of, e.g., broad-absorption line QSOs,
without carrying a major fraction of the energy of the AGN in kinetic
form.  In the NLS1s studied here, however, the UV absorption shows
only very low velocities, so the kinetic content of the gas is not
a problem.

\section{Spectral Energy Distributions (SEDs)}

Few near-simultaneous SEDs of NLS1s have been published.  With the
extremes of X-ray variability seen in some objects, 
simultaneity is a major concern.  Grupe et al. (1995) show an 
optical-to-X-ray SED of WPVS007.  
Our optical-to-UV SED is redder than theirs, but we have no 
corresponding X-ray data.  With the large number of optical telescopes 
available, it seems a shame not to obtain at least 
optical spectra simultaneous with future X-ray observations.

\end{document}